\documentclass{article}
\usepackage{./latex_resources/spconf}
\usepackage{amsmath,graphicx}

\usepackage[english]{babel}

\usepackage{ifpdf}

\usepackage{cite} 
\usepackage{url}
\usepackage{hyperref}
\usepackage{color}
\usepackage{pgf, tikz, pgfplots}
\usetikzlibrary{shapes, arrows, automata, plotmarks}
\usetikzlibrary{calc,hobby,decorations}
\usepackage{tikz-cd}

\usepackage{stfloats}

\usepackage{amsfonts, amssymb, amsthm}

\usepackage{subfloat}
\usepackage{enumerate}
\usepackage{multirow}
\usepackage{rotating}
\usepackage{subcaption}
\usepackage{needspace}

\input{./latex_resources/mySymbol.sty}
\input{./latex_resources/pennColors.sty}

\newtheorem{example}{\hspace{0pt}\bf Example}

\newtheorem{theorem}{\hspace{0pt}\bf Theorem}

\newtheorem{remark}{\hspace{0pt}\bf Remark}

\newtheorem{definition}{\hspace{0pt}\bf Definition}

\title{Quiver Signal Processing (QSP)}

\name{Alejandro Parada-Mayorga, Hans Riess, Alejandro Ribeiro,  and Robert Ghrist}
\address{University of Pennsylvania}
\begin{document}
\maketitle
\begin{abstract}
In this paper we state the basics for a signal processing framework on quiver representations. A quiver is a directed graph and a quiver representation is an assignment of vector spaces to the nodes of the graph and of linear maps between the vector spaces associated to the nodes. Leveraging the tools from representation theory, we propose a signal processing framework that allows us to handle heterogeneous multidimensional information in networks. We provide a set of examples where this framework provides a natural set of tools to understand apparently hidden structure in information.
We remark that the proposed framework states the basis for building graph neural networks where information can be processed and handled in alternative ways.
\end{abstract}
%




\section{Introduction}
\label{sec:intro}

Now ubiquitous for modeling complex systems, graphs serve as a popular canvas for information processing over data domains whose structure is not entirely known (e.g.~robots navigating an uncertain environment) or whose structure exhibits irregularity (e.g.~a proximity graph of a mobile sensor network).
Graphs prove useful in machine learning applications, in particular, graph (convolutional) neural networks (GNNs).
Numerical evidence \cite{gama_ggns} as well as stability properties \cite{gama_stability, parada_algnn} provide keen justification for deploying GNNs to model graph-like domains, not only to handle more generic representations of data but to obtain new insights about existing models.

Graphs are commonly described by matrices whose entries obey a sparsity pattern given by the connectivity between nodes in the graph. A \textit{signal} on a graph is a choice of vector whose components are defined on individual nodes of the graph i.e.~one component (scalar) of the signal per node, and a filter is a polynomial function of a matrix shift operator \cite{gama_stability} usually selected as the Lapalcian or the adjacency matrix. This rudimentary linear-aglebraic view of graph signal processing, while extensively and successfully studied \cite{gama_ggns}, is narrow in scope.
This framework is limited when it comes to modeling systems whose agents not only pass messages in a heterogenous environment, but also process heterogenous data types. Such data types may be vector-valued of varied or uniform dimension, set-valued \cite{puschel2018sets}, Boolean, or even lattice-valued \cite{riess2020tarski}. In this paper, we focus our attention on (finite dimensional) vector-valued data and information sharing (i.e.~message-passing) via linear transformations (as in \cite{hansen2020discourse}). 

Recent efforts have been made to model networks and information over them in more general ways. Cellular sheaves \cite{curry2014sheaves,shepard1985cellular} are a data structure for stitching together assignments of data to both nodes and edges of a graph. This powerful framework combined with insights from combinatorial Hodge theory \cite{eckmann1944hodge} has prompted interest in diffusion dynamics \cite{hansen2019spectral,riess2020tarski} (i.e.~a heat equation) and spectral properties \cite{hansen2019spectral} for cellular sheaves, and has led to applications in opinion dynamics \cite{hansen2020discourse}, distributed optimization \cite{hansen2019distributed}, and learning models for smooth graph signals \cite{hansen2019learning}.
This line of research makes copious use of sheaf Laplacians, both linear \cite{hansen2019spectral} and order-theoretic \cite{riess2020tarski}, corresponding to vector-valued and lattice-valued data respectively. In particular, the graph Laplacian, a common choice of shift operator in graph signal proccessing (GSP), can be viewed as the sheaf Laplacian of a particularly simple sheaf.

Parallel efforts, especially by P{\"u}schel et al, have been made at directly processing sets \cite{puschel2018sets} and meet/join lattices \cite{puschel2019lattices}. A convolutional neural information processing model was proposed for the former \cite{wendler2019powerset}, and an architecture employing the convolution of the later was recently used to classify multidimensional persistence modules \cite{riess2020persistence}.
Others have contributed category-theoretic perspectives on networks, for example, making use of convenient mathematical objects (of an algebraic nature) such as operads \cite{baez2020network} or groupoids \cite{dehaan2020natural}.

While sheaves have promise for modeling information systems, they require ``hidden'' data assigned to edges. In some instances this is inefficient or even unnatural as data must be \textit{pulled back} from all edges incident to a node before it may be processed by a node. 
Furthermore, cellular sheaves are not a natural object for studying directed graphs, although some have bravely tried to tame sheaves for the directed setting \cite{krishnan2014flow}. Generalized persistence modules \cite{bubenik2013metrics}, which encompass both cellular sheaves and persistent homology \cite{ghrist2008barcodes}, are adaptable to directed graphs; they are slightly more general than the quiver representations studied here.

In this work we propose a new signal processing framework relying on the representation theory of quivers.
Quiver representation theory has a rich history \cite{repthysmbook,derksen2017introduction}, and while it has impacted many other areas of mathematics \cite{repthysmbook,derksen2017introduction}, it has not yet seen the spotlight in engineering.

A quiver is a jawn\footnote{Philadelphian vernacular for an ambiguous \textit{thing} or \textit{object}.} containing arrows and nodes (i.e.~a directed multigraph). A quiver representation is an assignment of a vector space to each node and an assignment of a linear map associated to each arrow. A quiver representation provides a natural environment to handle heterogeneous information as vector spaces associated to nodes can be non-isomorphic i.e.~have different dimensions. Leveraging the fundamentals of representation theory, we can extend the notions of filtering and Fourier decomposition to quiver representations. We provide examples along the way including an example from arising from persistent homology and one arising from a system of heterogenous autonomous agents. We highlight that our proposed approach to filtering has the potential to extend GNNs to a more general setting of quiver neural networks (QNNs).

The paper is organized as follows. In Section 2 we introduce the basics of quiver representation theory while in Section 3 we present a signal processing framework for signals on quivers, and in Section 4 we present some conclusions.




\section{Quiver Representations}
\label{sec:quivrep}


\begin{figure}
	\centering
\includegraphics[width=0.18\textwidth]{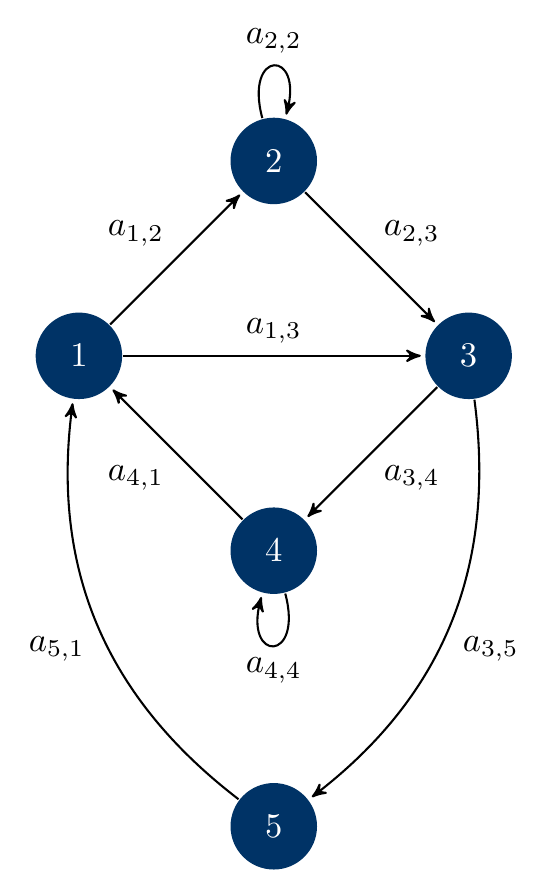}
\includegraphics[width=0.18\textwidth]{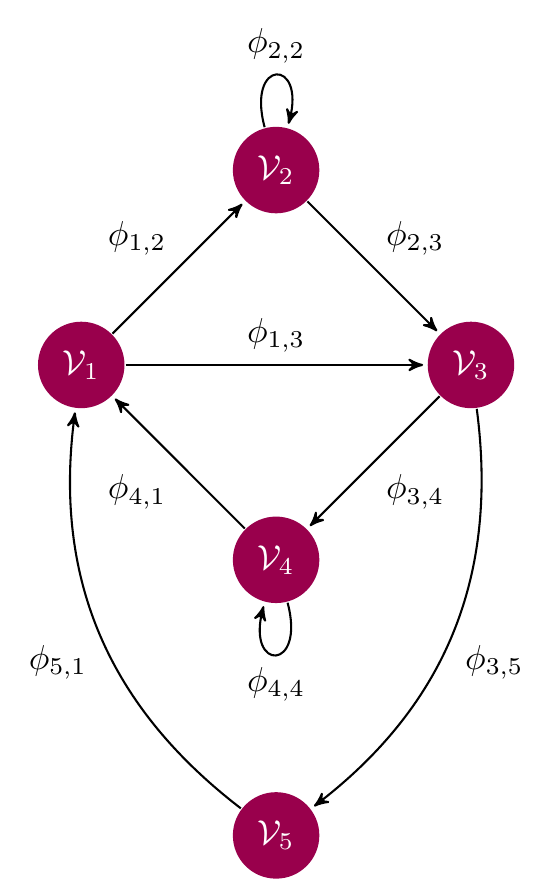}
\caption{A quiver $Q$ (left) and one  particular representation (right) of $Q$. $\mathcal{V}_{i}$ is a vector space associated to the node $i$ of the quiver while $\phi_{i,j}$ is a linear map between the spaces $\mathcal{V}_{i}$ and $\mathcal{V}_{j}$.}
\label{fig:quiver1}
\end{figure}

A quiver is a directed graph, permitting loops and multiple arrows between the same  set of nodes. Concretely, a quiver is described by the following data.

\begin{definition}\label{def:quiver}
A \textbf{quiver} $Q$ is a quadruple $Q=(Q_{0},Q_{1},h,t)$ consisting of sets $Q_0$ and $Q_1$, and maps $h,t: Q_1 \rightarrow Q_0.$
\end{definition}
Some terminology: $Q_0$ are the \textbf{nodes} of $Q$ and $Q_1$ are the \textbf{arrows}.
For an arrow $a \in Q_1$, we say $h(a)$ is the \textbf{head} of $a$ and $t(a)$ is the \textbf{tail} of $a$.


\begin{example}\label{example:1}\normalfont
Let us consider the quiver depicted in Fig.~\ref{fig:quiver1} (left).
Let $Q_{0}=\{1,2,3,4,5\}, $
and $Q_{1}=\{a_{1,2}, a_{2,3},a_{2,2},a_{3,4},\allowbreak a_{3,5},a_{4,4},a_{4,1},a_{5,1}\}.$
The maps $h$ and $t$ are given by $h(a_{i,j})=j$ and $t(a_{i,j})=i$.
\end{example}


Quivers alone do not offer anything new to the GSP literature. However, the notion of a representation of a quiver provides the needed formal components used to associate information to a quiver in different ways.


\begin{definition}
A (finite-dimensional) \textbf{representation} $\boldsymbol{\pi}$ of a quiver $Q = (Q_0, Q_1, h, t)$ is an assignment of (finite-dimensional) vector spaces to nodes: $\boldsymbol{\pi}: i \in Q_0 \mapsto \mathcal{V}_i,$
and an assignment of linear maps to arrows: $\boldsymbol{\pi}: a \in Q_1 \mapsto
\left( \phi_{t(a), h(a)}: \mathcal{V}_{t(a)} \rightarrow \mathcal{V}_{h(a)}\right).$
\end{definition}

%



\begin{example}[robotics]\normalfont
Quiver representations offer a formal tool to analyze and understand heterogeneous large-scale distributed robotic systems~\cite{robotappl1}. In particular, we can define an arbitrary graph where each node is a robot in the system. To each node in the graph, we associate a vector space that decomposes as $\ccalV_{\text{sense}}\oplus\ccalV_{\text{task}}$ to represent the type of information that each robot can sense $\ccalV_{\text{sense}}$,  and the configuration space of the task to be performed by the robot $\ccalV_{\text{task}}$. Message-passing between robots is given by linear operators that synthesize both sensing and task-related observations, according the design of the control system. This setup is natural when considering the classification of robots in several \textit{species} in the sense defined in~\cite{robotappl1}. 
\end{example}



\begin{example}[topological data analysis]\normalfont
Quiver representations arise naturally when studying the homology of a filtered space~ \cite{oudot2017persistence,ghrist2014elementary}. In laymans terms, homology is a vector space associated to a topological space that counts the number of connected components, loops, voids, and higher dimensional features. A filtered space is a sequence of inclusions of topological spaces
\[ \mathbb{X}_0 \subseteq \mathbb{X}_1 \subseteq \cdots \subseteq \mathbb{X}_n.\]
Examples are plentiful, including graphs filtered by edge weights or the Rips complex \cite{ghrist2008barcodes} of a point cloud in $\mathbb{R}^d$ filtered by radii.
Homology induces linear transformations between successive spaces in a filtration.

The setup above can be succintly described as a certain representation of the quiver $Q:~\circ \longrightarrow \circ \longrightarrow \cdots \longrightarrow \circ$.
The representation of interest is given by homology (a vector space) of spaces in the filtration and the (linear) maps induced by the inclusion of spaces,
\[ H_k(\mathbb{X}_0) \rightarrow H_k(\mathbb{X}_1) \rightarrow \cdots \rightarrow H_k(\mathbb{X}_n).\]
The observation that the homology of a filtration is in fact a quiver representation has lead to a guaranteed compact description of a filtration called a \textit{barcode} \cite{ghrist2008barcodes}. Work extending such quivers to the entire real line has lead to similar pleasant decompositions into irreducibles~\cite{crawleyboevey2012decomposition}.
\end{example}


The theory of quiver representations  is naturally embedded in the context of the representation theory of algebras~\cite{repthysmbook,derksen2017introduction}. Therefore, it is possible to talk about irreducible representations, indecomposable representations and so on. In what follows, we proceed to introduce these notions for quivers, which will be later used to define a signal processing framework. We start with a notion that will allow us to compare representations of a quiver.


\begin{definition}
Let $\boldsymbol{\pi}$ and $\boldsymbol{\rho}$ be representations of a quiver $Q$. An \textbf{intertwining map} $T: \boldsymbol{\pi} \rightarrow \boldsymbol{\rho}$ is a family of linear transformations $\{ T_i: \boldsymbol{\pi}(i) \rightarrow \boldsymbol{\rho}(i)~\vert~i \in Q_0\}$ such that for every arrow $a \in Q_1$,
\begin{center}
\begin{tikzcd}
\boldsymbol{\pi}(t(a)) \arrow[r, "\boldsymbol{\pi}(a)"] \arrow[d, "T_{t(a)}"]
& \boldsymbol{\pi}(h(a)) \arrow[d, "T_{h(a)}" ] \\
\boldsymbol{\rho}(t(a)) \arrow[r, "\boldsymbol{\rho}(a)" ]
& \boldsymbol{\rho}(h(a)).
\end{tikzcd}
\end{center}
We say that $T$ is an isomorphism (of representations) if  the $T_{i}$ are invertible for all $i\in Q_{0}$.
\end{definition}


We will use the symbol $\cong$ to denote isomorphism between representations. Now, for any two representations $\boldsymbol{\pi}_{1}$ and $\boldsymbol{\pi}_{2}$ of a quiver $Q$, their \textbf{direct sum} is given by $(\boldsymbol{\pi}_{1} \oplus \boldsymbol{\pi}_{2})(i)=\boldsymbol{\pi}_{1}(i)\oplus\boldsymbol{\pi}_{2}(i)$ whether $i$ is in $Q_{0}$ or $Q_{1}$~\cite{derksen2017introduction}. We will call a representation $\boldsymbol{\pi}$ \textbf{trivial} if $\boldsymbol{\pi}(x)=\mathbf{0}$ for all $x\in Q_{0}$.

Now, we are ready to introduce the notion of an indecomposable and an irreducible representation.


\begin{definition}
A nontrivial representation $\boldsymbol{\pi}$ of a quiver $Q$ is \textbf{decomposable} if $\boldsymbol{\pi}$ is isomorphic to $(\boldsymbol{\pi}_{1}\oplus\boldsymbol{\pi}_{2})$ for some nontrivial representations 
$\boldsymbol{\pi}_{1}, \boldsymbol{\pi}_{2}$ of $Q$. A representation that is not decomposable is called \textbf{indecomposable}.
\end{definition}
Recall a \textbf{subrepresentation} $\boldsymbol{\theta}$ of $\boldsymbol{\pi}$ is a restriction of $\boldsymbol{\theta}$ to subspaces $\mathcal{W}_i$ of $\mathcal{V}_i,~i \in Q_0$
\begin{definition}
A nontrivial representation $\boldsymbol{\pi}$ is \textbf{irreducible} if for every subrepresentation $\boldsymbol{\theta}$ of $\boldsymbol{\pi}$, $\boldsymbol{\theta}$ is trivial or equal to $\boldsymbol{\pi}$.
\end{definition}


Representations of quivers are equivalent to representations of certain associative algebras.
In particular, the traditional notions of decomposability and irreducibiity can be studied in the algebra setting \cite{derksen2017introduction,repthysmbook}.

A \textbf{path} $\gamma$ in a quiver $Q=(Q_{0},Q_{1},h,t)$ of length $\ell\geq 1$ is a sequence $\gamma=a_{\ell}a_{\ell-1}\ldots a_{1}$ such that $a_i \in Q_{1}$ or all $i$ and $t(a_{i+1})=h(a_{i})$ for $i=1,\ldots,\ell-1$ with $h(\gamma)=h(a_{\ell})$ and $t(\gamma)=t(a_{1})$~\cite{derksen2017introduction}. By convention, there is a \textbf{trivial path} $e_{i}$ of length zero can be defined by $h(e_{i})=t(e_{i})=i$ for every $i \in Q_0$. Let $\mathrm{Path}(Q)$ denote the set of paths of a quiver $Q$.


\begin{definition}
Let $k$ be a field. The \textbf{path algebra} of a quiver $Q = (Q_0, Q_1, h, t)$, denoted $k Q$, is the free vector space with basis $\langle \mathrm{Path}(Q) \rangle$ and a product on basis elements,
\[
\gamma_2 \cdot \gamma_1 = 
\begin{cases}
\gamma_2 \gamma_1 & t(\gamma_2) = h(\gamma_1) \\
0 				 & t(\gamma_2) \neq h(\gamma_1)
\end{cases}.
\]
(This product extends to a product on $k Q$ by linearity.)
\end{definition}


\begin{remark}\normalfont
If $Q$ is a finite quiver, $kQ$ is unital with unit $1 = \sum_{i \in Q_0} e_i$.
This is seen by observing $e_i \cdot e_j = \delta_{i j} e_i$ (where $\delta_{ij} = 1$ iff $i = j$ and $0$ otherwise). Additionally, notice that $kQ$ is \textit{generated} by $e_{i}$ and the arrows $a_{i,j}\in Q_{1}$.
\end{remark}
Recall, a \textbf{representation of an algebra} $\mathcal{A}$ is given by a pair $(\ccalM,\rho)$, where $\ccalM$ is a vector space and $\rho$ is a so-called homomorphism $ \rho: \mathcal{A} \rightarrow \mathrm{End}(\mathcal{M})$,
where $\mathrm{End}(\mathcal{M})$ is the algebra of linear maps from a vector space $\mathcal{M}$ to itself. Notice that $\rho$ is a homomorphism if $\rho$ is a linear map, $\rho$ respects products $\rho(a \cdot b) = \rho(a) \cdot \rho(b),$ and and $\rho(1) = I$.

The following result states formally the correspondence between the representations of a quiver and the representations of its path algebra.


\begin{theorem}[\cite{derksen2017introduction}]\label{theorem:equivrep}
There is a bijection
\[
\{ \text{Repesentations of } Q \} \leftrightarrow \{ \text{Representations of } k Q \}
\]
(In fact, this bijection is an equivalence of categories.)
\end{theorem}

One direction of Theorem~\ref{theorem:equivrep} implies that a given quiver representation $\boldsymbol{\pi}$ can be transformed into an equivalent representation of $kQ$, which consists of the data of a vector space $\ccalM$ and a homomorphism $\rho$. Indeed $\ccalM=\oplus_{i}\ccalV_{i}$ where $\ccalV_{i}=\boldsymbol{\pi}(i)$ and $i\in Q_{0}$. We also point out that because the path algebra $kQ$ is \textit{generated} by the elements in $\mathrm{Path}(Q)$, we can describe the action of $\rho$ in terms of $\rho(p)$ for all $p\in \mathrm{Path}(Q)$. In particular, the action of an element $p\in \mathrm{Path}(Q)$ on an element $\bbx\in\ccalM$ is given by $\bby=\rho(p)\bbx$ where
\begin{equation*}
\bby(j)= 
              \left\lbrace
                    \begin{array}{cc}
                    \boldsymbol{\pi}(p)\bbx(i)  &   \text{if}\quad t(p)=i, h(p)=j\\
                    0            &   \text{otherwise},  
                    \end{array}
              \right.
\end{equation*}
where $\boldsymbol{\pi}(p)$ for the path $p=a_{\ell}a_{\ell-1}\ldots a_{1}$ is given naturally by the composition map $\boldsymbol{\pi}(p)=\boldsymbol{\pi}(a_{\ell})\boldsymbol{\pi}(a_{\ell-1})\ldots\boldsymbol{\pi}(a_{1})$.





\section{Signal Processing on Quiver Representations}
\label{sec:spquiverep}

We now proceed to develop a theory of signal processing on quiver representations under the umbrella of algebraic signal processing as defined in~\cite{algSP0}.


\begin{definition}\label{def:quiversig}
Let $Q=(Q_{0},Q_{1},h,t)$ a quiver and $\boldsymbol{\pi}$ a representation of $Q$. Then a \textbf{signal} $\mathbf{x}$ on $\boldsymbol{\pi}$ is an element $\bbx \in \bigoplus_{i \in Q_0} \boldsymbol{\pi}(i).$
%
%
Additionally, the elements in  $kQ$ are called \textbf{algebraic filters} while their images in $\mathrm{End}(\mathcal{M})$ via the homomorphism $\rho$ associated to the representation of $kQ$ are called \textbf{quiver filters}.
\end{definition}
%

%
%
%
%
Then, as stated in~\cite{algSP0}, the filtering of a signal $\bbx$ is given by $\rho(c)\bbx$ where $c\in kQ$. Notice that this is the \textbf{convolution} between the signal $\bbx$ and $\rho(c)$. In order clarify this notion of filtering, we present the following example. 


\begin{example}[Filtering]\normalfont
Taking into account the quiver depicted in Fig.~\ref{fig:quiver1} (left) and its representation in Fig.~\ref{fig:quiver1} (right) we have an example of filtering considering the algebraic filter $c=a_{5,1}a_{3,5}+ a_{1,2}a_{4,1}a_{3,4}+a_{1,3}a_{2,3}$. Then, when filtering a signal $\bbx \in \oplus_{i \in Q_0} \ccalV_{i}$, we have
\begin{equation*}
\bby=\rho(c)\bbx=\rho\left(a_{5,1}a_{3,5}+ a_{1,2}a_{4,1}a_{3,4}+a_{1,3}a_{2,3}\right)\bbx.
\end{equation*}
Now, taking into account the linearity of $\rho$ it follows that
\begin{equation*}
\bby=\rho(a_{5,1}a_{3,5})\bbx+\rho(a_{1,2}a_{4,1}a_{3,4})\bbx+\rho(a_{1,3}a_{2,3})\bbx,
\end{equation*}
and as a consequence of the action of $\rho$ on elements of $\ccalM$ we have $\bby(1)=\phi_{5,1}\phi_{3,5}\bbx(3)$, $\bby(2)=\phi_{1,2}\phi_{4,1}\phi_{3,4}\bbx(3)$ and $\bby(\ell)=\mathbf{0}$ for $\ell=3,4,5$.
\end{example}


Taking into account that $kQ$ is generated by the elements in $\mathrm{Path}(Q)$, their image under the homomorphism in a representation are useful to represent quiver filters. Due to their importance in the characterization of an algebraic signal model, we provide a formal name for them as follows.


\begin{definition}
Let $Q=(Q_{0},Q_{1},h,t)$ a quiver and $\boldsymbol{\pi}$ a representation of $Q$. Let $\rho$ the homomorphism induced by $\boldsymbol{\pi}$ in a representation of the path algebra $kQ$. Then, the operator $\rho(p)$ for any $p\in\mathrm{Path}(Q)$ is called a \textbf{shift operator}.
\end{definition}

Now we turn our attention to the frequency domain. As pointed out in~\cite{algSP0}, a notion of Fourier decomposition can be built from irreducible subrepresentations of the algebra in the signal model. Following this line, we proceed to introduce the notion of Fourier decompositions using the path algebra $kQ$, the vector space $\ccalM= \bigoplus_{i \in Q_0} \boldsymbol{\pi}(i)$ and the homomorphism $\rho$ as described above.


\begin{definition}[Fourier Decomposition]\label{def:foudecomp}
Let $Q=(Q_{0},Q_{1},h,t)$ a quiver, $\boldsymbol{\pi}$ a representation of $Q$ and $(\ccalM,\rho)$ the representation of $kQ$ induced by $\boldsymbol{\pi}$ with $\ccalM= \bigoplus_{i \in Q_0} \boldsymbol{\pi}(i)$. Then, we say that there is a spectral or Fourier decomposition on $\boldsymbol{\pi}$ if
\begin{equation}
(\mathcal{M},\rho)\cong\bigoplus_{(\mathcal{U}_{i},\phi_{i})\in\text{Irr}\{\mathcal{A}\}}(\mathcal{U}_{i},\phi_{i})^{\oplus m(\mathcal{U}_{i},\mathcal{M})}
 \label{eq:foudecomp1}
\end{equation}
where the $(\mathcal{U}_{i},\phi_{i})$ are irreducible subrepresentations of $(\mathcal{M},\rho)$ and $m(\ccalU_{i},\ccalM)$ is the number of irreducible subrepresentations of $\ccalM$ isomorphic to $\ccalU_{i}$. Any signal $\mathbf{x}\in\mathcal{M}$ can be therefore represented by the map $\Delta$ given by 
\begin{equation}
\begin{matrix} 
\Delta: & \mathcal{M} \to \bigoplus_{(\mathcal{U}_{i},\phi_{i})\in\text{Irr}\{\mathcal{A}\}}(\mathcal{U}_{i},\phi_{i})^{\oplus m(\mathcal{U}_{i},\mathcal{M})} \\ & \mathbf{x}\mapsto \hat{\mathbf{x}}
 \end{matrix}
 \label{eq:foudecomp2}
\end{equation}
known as the Fourier decomposition of $\mathbf{x}$, and the projection of $\hat{\mathbf{x}}$ in each $\mathcal{U}_{i}$ are the Fourier components.
\end{definition}


Notice that $\Delta$ is an intertwining map, therefore $\Delta(\rho(c)\bbx)=\rho(c)(\Delta(\bbx))$. Using this fact, a spectral representation of the operator $\rho(c)$ can be derived. It is worth pointing out that the maps $\phi_{i}$ indicating the spectral components of the Fourier decomposition are not necessarily scalar like in the case of commutative algebras~\cite{algSP0,repthybigbook}.

Another tool we can consider for the decomposition of a quiver signal is associated to the indecomposable representations on a quiver, which are not necessarily irreducible but they constitute an important building block in quiver theory~\cite{derksen2017introduction}. In particular, we highlight that the indecomposable representations of a quiver are determined by the topology of the quiver considering a graph representation where the arrows are replaced by undirected edges~\cite{repthysmbook,derksen2017introduction}. We formally name these decompositions in the following definition, as they can highlight properties of a signal not reflected in the Fourier decomposition.


\begin{definition}
Let $Q=(Q_{0},Q_{1},h,t)$ be a quiver, and consider $\boldsymbol{\theta}_{i}$ the indecomposable representations of $Q$. Then, we say that $\boldsymbol{\pi}$ has a \textbf{basic decomposition} if 
\begin{equation}
\boldsymbol{\pi}\simeq r_{1}\boldsymbol{\theta}_{1}\oplus r_{2}\boldsymbol{\theta}_{2}\oplus\cdots\oplus r_{k}\boldsymbol{\theta}_{k}
\end{equation}
where $r_{i}\boldsymbol{\theta}_{i}$ represent the direct sum of $r_{i}$ copies of $\boldsymbol{\theta}_{i}$. 
\end{definition}


\begin{example}[Basic decomposition]\normalfont
Let us consider the quiver $Q$: 
\begin{tikzcd}
\circ \arrow[r,] & \circ  
\end{tikzcd}
which consists only of two nodes and one arrow, and consider the representation $\boldsymbol{\pi}$ of $Q$ given by
\begin{tikzcd}
\mathbb{C}^{r+s} \arrow[r,  "\phi_{1,2}" ] & \mathbb{C}^{r+t},
\end{tikzcd}
where
\begin{equation*}
\phi_{1,2}
=
\begin{bmatrix}
\mathbf{I}_{r\times r}  & \mathbf{0}_{r\times s}\\
\mathbf{0}_{t\times r}  & \mathbf{0}_{t\times s}
\end{bmatrix}.
\end{equation*}
The three indecomposable representations of Q are~\cite{repthysmbook,derksen2017introduction}: \\
(Rep1) 
\begin{tikzcd}
\mathbb{C}\arrow[r, "1"] & \mathbb{C}  
\end{tikzcd}, 
(Rep2)
\begin{tikzcd}
\mathbf{0}\arrow[r, "0"] & \mathbb{C}  
\end{tikzcd}, and \\
(Rep3)
\begin{tikzcd}
\mathbb{C}\arrow[r, "0"] & \mathbf{0}
\end{tikzcd}.
We can see that $\boldsymbol{\pi}$ can expressed as a direct sum of $r$ copies of (Rep1), $s$ copies of (Rep3) and $t$ copies of (Rep2).
\end{example}


%
%
%
%
%




\section{Conclusions}\label{sec:conclusions}

We have lain the groundwork for a signal processing framework on quiver representations. Exploiting algebraic signal processing theory and the representation theory of algebras, we introduced the notions of signals, filters, spectral representations, and basic decompositions. We provided basic examples to familiarize the reader with quiver representations, and we suggested application areas (robotics and computational topology). We have shown how an existing theoretical machinery from representation theory can be leveraged to manipulate and extract information in ways not explored before. This signal processing framework provides a new tool for handling heterogenous data distributed across networks, opening the door for new neural convolutional architectures. As an additional future research direction, we intend to explore physical applications of our framework in the prequel.

%
%


\bibliographystyle{IEEEbib}
\bibliography{biblio}

\end{document}